# Bulk Fermi surface of the layered superconductor TaSe$_3$ with three-dimensional strong topological insulator state

Wei Xia[1,2,3,†], Xianbiao Shi[4,5,†], Yong Zhang[6,3,7,†], Hao Su[1,2,3], Qin Wang[8,15], Linchao Ding[10], Leiming Chen[11], Xia Wang[1,12], Zhiqiang Zou[1,12], Na Yu[1,12], Li Pi[13], Yufeng Hao[8,9], Bin Li[14], Zengwei Zhu[10], Weiwei Zhao[4,5,*], Xufeng Kou[6,*], Yanfeng Guo[1,*]

[1] School of Physical Science and Technology, ShanghaiTech University, Shanghai 201210, China
[2] Shanghai Institute of Optics and Fine Mechanics, Chinese Academy of Sciences, Shanghai 201800, China
[3] University of Chinese Academy of Sciences, Beijing 100049, China
[4] State Key Laboratory of Advanced Welding & Joining and Flexible Printed Electronics Technology Center, Harbin Institute of Technology, Shenzhen 518055, China
[5] Key Laboratory of Micro-systems and Micro-structures Manufacturing of Ministry of Education, Harbin Institute of Technology, Harbin 150001, China
[6] School of Information Science and Technology, ShanghaiTech University, Shanghai 201210, China
[7] Shanghai Institute of Microsystem and Information Technology, Chinese Academy of Sciences
[8] College of Engineering and Applied Sciences, Nanjing University, Nanjing 210093, China
[9] Haian Institute of New Technology, Nanjing University, Haian, 226600, China
[10] Wuhan National High Magnetic Field Center and School of Physics, Huazhong University of Science and Technology, Wuhan 430074, China
[11] School of materials science and engineering, Henan key laboratory of aeronautic materials and application technology, Zhengzhou University of Aeronautics, Zhengzhou, Henan, 450046
[12] Analytical Instrumentation Center, School of Physical Science and Technology, ShanghaiTech University, Shanghai 201210, China
[13] Anhui Province Key Laboratory of Condensed Matter Physics at Extreme Conditions, High Magnetic Field Laboratory of the Chinese Academy of Sciences, Hefei, China
[14] Information Research Center, Nanjing University of Post and Telecommunication, Nanjing, 210023, China
[15] School of Physics and Microelectronics, Zhengzhou University, Zhengzhou, Henan, 450001

High magnetic field transport measurements and *ab initio* calculations on the layered superconductor TaSe$_3$ have provided compelling evidences for the existence of a three-dimensional strong topological insulator state. Longitudinal magnetotransport measurements up to ~ 33 T unveiled striking Shubnikov-de Hass oscillations with two fundamental frequencies at 100 T and 175 T corresponding to a nontrivial electron Fermi pocket at the B point and a nontrivial hole Fermi pocket at the Γ point respectively in the Brillouin zone. However, calculations revealed one more electron pocket at the B point, which was not detected by the magnetotransport measurements, presumably due to the limited carrier momentum relaxation time. Angle dependent quantum oscillations by rotating the sample with respect to the magnetic field revealed clear changes in the two fundamental frequencies, indicating anisotropic electronic Fermi pockets. The *ab initio* calculations gave the topological $Z_2$ invariants of (1; 100) and revealed a single Dirac cone on the (1 0 -1) surface at the $\bar{X}$ point with helical spin texture at a constant-energy contour, suggesting a strong topological insulator state. The results demonstrate $TaSe_3$ an excellent platform to study the interplay between topological phase and superconductivity and a promising system for the exploration of topological superconductivity.

[†]The authors contributed equally to this work.

[*]Corresponding authors:
wzhao@hit.edu.cn,
kouxf@shanghaitech.edu.cn
guoyf@shanghaitech.edu.cn.

## I. INTRODUCTION

A distinct feature of Majorana Fermion, which could be realized in solids as quasiparticle, is that it is its own antiparticle and the bound state obeys non-Abelian statistics, thus making it potential for the use in low decoherence topological quantum computation [1-6]. In addition to many proposed systems such as the superfluid $^3$He [7] and the $v = 5/2$ fractional quantum Hall system [8], etc., the topological superconductors (TSs) are also thought to host Majorana Fermion because the interplay of nontrivial band topology and superconductivity is expected to give rise to topological superconductivity with spin-momentum locking of topological surface states, which is capable of creating a solid-state environment for the Majorana Fermion [9-11]. In the theoretical scheme, the Majorana Fermion behaving as Majorana zero modes occurs in general in vortices and on edges of effectively spinless superconducting systems with odd-parity triplet pairing symmetry [4-8], that is, *p*-wave pairing symmetry in one dimension (1D) and $p_x \pm ip_y$ pairing symmetry in two dimensions (2D). The $Sr_2RuO_4$ [12, 13] and $Cu_xBi_2Se_3$ superconductors, [14, 15] which have been proposed to possess *p*-wave triplet pairing ground state, have therefore gained considerable attentions since it is known that *p*-wave superconductors are very sparse [11]. However, a consensus on the *p*-wave pairing symmetry in the two superconductors has not been reached yet. An alternative equivalent $p + ip$ pairing superconductor [4,9], i.e. a heterostructure constructed by a topological insulator (TI) and an *s*-wave conventional superconductor [16-18], was experimentally verified to host topological superconductivity induced by the proximity effect in the spin-helical topological surface states [4,9]. A single material with equivalent $p + ip$ type superconductivity, which is free from the disturbance of the interface physics as that in a heterostructure, is preferable for exploring Majorana Fermion, thus guiding tremendous efforts to examining the nontrivial topological states in known superconductors. Majorana zero mode was argued to be observed in some superconductors such as $FeTe_{0.55}Se_{0.45}$ ($T_c$ = 14.5 K) [19-22], $(Li_{0.84}Fe_{0.16})OHFeSe$ [23,24], $CaKFe_4As_4$ ($T_c$ = 35 K) [25], and a 2M-phase $WS_2$ ($T_c$ = 8.8 K) [26], suggesting that intrinsic TSs are actually suitable systems for the exploration.

The layered transition-metal trichalcogenides $TrX_3$ ($Tr$ = Nb, Ta; $X$ = S, Se), with the schematic crystal structure of the representative $TaSe_3$ shown in Fig. 1(a), are a family of superconductors that have not been well studied because of the rather low $T_c$ of ~ 2 K. Previous attentions on them were primarily paid on the possible interplay between charge-density-wave (CDW) states and superconductivity [27-30]. As an exception, the trigonal prismatic $TaSe_3$ has gained less interest since its superconducting properties were observed decades ago [27], due to the absence of CDW transition. However, studies exposed high current-carrying capacity of $TaSe_3$ [31], implying that this material has potential use as an interconnector in electronic devices. Besides, it was found that quasi-1D nanowires of $TaSe_3$ had lower levels of normalized noise spectral density [32,33], thus offering potential for downscaled local interconnect applications. However, knowledge on the electronic band structure of $TaSe_3$ is still very limited, which therefore motivated us to investigate this ever overlooked superconductor by using magnetotransport measurements and *ab initio* calculations. Interestingly, our work revealed a three-dimensional (3D) strong topological insulator (TI) phase with $Z_2$ invariants (1; 100) and a single Dirac cone on the $(10\bar{1})$ surface at the $\bar{X}$ point with helical spin texture at a constant-energy contour. The result is nicely consistent with the results of a very recent theoretical work [34].

## II. EXPERIMENTAL

The TaSe$_3$ crystals were grown by using the self-flux method. Starting materials of Ta powder (99.9%, Aladdin) and Se (99.999%, Aladdin) granules were mixed in a molar ratio of 1 : 6 and placed into an alumina crucible. The assembly was heated in a furnace up to 750 °C within 9 hrs, kept at the temperature for 20 hrs, and then slowly cooled down to 450 °C at a temperature decreasing rate of 1.5 °C/h. The excess Se was removed at this temperature by quickly placing the assembly into a high-speed centrifuge. Then, the quartz tube was cooled to room temperature in air. The shining quasi 1D crystals with a typical dimension of 2.4 × 0.1 × 0.06 mm$^3$ were obtained. The typical picture of a crystal is presented in the inset of Fig. 1(b).

The phase and quality examinations of TaSe$_3$ were performed on a Bruker D8 single crystal X-ray diffractometer (SXRD) with Mo $K_{\alpha 1}$ ($\lambda$ = 0.71073 Å) at 298 K. The diffraction pattern can be well indexed on the basis of a trigonal structure with the lattice parameters $a$ = 9.82160 Å, $b$ = 3.49520 Å, $c$ = 10.39720 Å, $\alpha$ = 90 °, $\beta$ = 106.3442 ° and $\gamma$ = 90 °, in the space group $P2_1/m$ (No. 11), consistent with that reported previously [29]. The schematic crystal structure in Fig. 1(a) is drawn based on the refinement results from the SXRD data. Fig. 1(c) shows the average compositions derived from a typical energy dispersive X-ray spectrum (EDS) measured at more than ten different points on the crystal, revealing good stoichiometry with the atomic ratio of Ta : Se = 1 : 3. The picture in Fig. 1(c) shows the crystal used for EDS measurements. The perfect reciprocal space lattice without any other miscellaneous points, seen in Figs. 1(d) - (f), indicates pure phase and high quality of the crystal. The magnetotransport measurements were carried out using a standard four-wire method at both High Magnetic Field Laboratory, Chinese Academy of Sciences and Wuhan National High magnetic field center.

First-principles methods based on the density functional theory (DFT) encoded in the Vienna *ab initio* Simulation Package (VASP) [35-37] was used. For all calculations, the cutoff energy for the plane-wave basis was set to 500 eV, the Brillouin zone (BZ) sampling was done with a Γ-centered Monkhorst-Pack k-point mesh of size 6 × 15 × 5, and the total energy difference criterion was defined as $10^{-8}$ eV for self-consistent convergence. The modified Becke-Johnson (mBJ) [38,39] potential at the meta-GGA level was adopted to obtain the accurate band inversion strength and the band order. The maximally localized Wannier functions (MLWFs) [40,41] for Ta-d and Se-p orbitals were constructed to determine the hopping parameter values for tight-binding model. The Wannier Tools package [42], which works in the tight-binding framework was used to compute the surface spectrum including the surface density of states and the Fermi arcs based on the iterative Green's function method [43].

## III. RESULTS AND DISCUSSION

The temperature dependence of longitudinal resistivity $\rho_{xx}$ was measured with the magnetic field $B$ perpendicular to the (1 0 -1) plane and the electrical current $I$ along the b-axis. The $\rho_{xx}$ at $B = 0$ T is presented in Fig. 1(b), which displays metallic conduction with decreasing the temperature, yielding a residual resistance ratio (*RRR*) $\rho_{xx}(300\ K)/\rho_{xx}(2\ K)$ of approximately 26. With the measured temperature limit to 1.8 K, no superconducting transition is observed, consistent with the fact that the transition is only around 2 K. When $B$ is applied and gradually increased, $\rho_{xx}$ is enhanced to be somewhat insulating with a plateau behavior at low temperature, which is commonly observed in many topological metals/semimetals [44-47]. When $T > 100$ K, the increase of $B$ whereas slightly reduces the $\rho_{xx}$, which might be related to the gentle change of Fermi surface (FS).

Thermal evolution of the *B* dependent magnetoresistance (MR), which is defined as MR = [ρ(B) − ρ(0)]/ρ(0) × 100% where ρ(B) and ρ(0) represent the resistivity with and without *B*, respectively, measured in the temperature range of 1.8 − 15 K, is presented in Fig. 2(a) by the main panel. The MR shows a crossover from a quadratic-like evolution to linear change with respect to *B*, reaching a large value of ~2.8 ×10$^3$% at 33 T and 1.8 K without showing any sign of saturation, implying that the MR at large *B* is presumably dominated by the quantum effect. Interestingly, clear quantum oscillations in the MR is observable when *B* > 12 T below 15 K. To our best knowledge, this behavior has not been observed before in TaSe$_3$.

To trace the origin of the quantum oscillations in the MR, a detailed analysis was conducted. After a careful subtraction of the smooth background of the MR, striking Shubnikov-de Hass (SdH) oscillations are visible. The SdH oscillations at different temperatures from 1.8 to 15 K against the reciprocal magnetic field 1/*B* are plotted in Fig. 2(b), which could be well described by the Lifshitz-Kosevich (L-K) formula [48]:

$$\Delta\rho_{xx} \propto R_T R_D \cos\left[2\pi\left(\frac{F}{B} + \varphi\right)\right],$$

where $R_T = 2\pi^2 k_B T/\hbar\omega_c / \sinh(2\pi^2 k_B T/\hbar\omega_c)$, $R_D = \exp(-2\pi^2 k_B T_D/\hbar\omega_c)$, $k_B$ is the Boltzmann constant, $\hbar$ is the Planck's constant, *F* is the frequency of oscillation, *φ* is the phase shift, $\omega_c = eB/m^*$ is the cyclotron frequency with *m*\* denoting the effective cyclotron mass, $T_D$ is the Dingle temperature defined by $T_D = \hbar/2\pi k_B \tau_Q$ with $\tau_Q$ being the quantum scattering lifetime. The fast Fourier transform (FFT) spectra of the SdH oscillations, depicted in Fig. 2(c), disclose four frequencies, $F_\alpha$ = 100 T, $F_\beta$ = 175 T, $F_{\alpha+\beta}$ = 275 T and $F_{2\beta}$ = 350 T. However, it is apparent that only two of them are fundamental frequencies because $F_{\alpha+\beta} = F_\alpha + F_\beta$ and $F_{2\beta} = 2 F_\beta$, indicating that the $F_{\alpha+\beta}$ and $F_{2\beta}$ are primarily arising from the combination of $F_\alpha$ and $F_\beta$. This fact indicates that at least two pockets across the Fermi level $E_F$, which however is somewhat different from the theoretical calculations that suggest three Fermi pockets across $E_F$. The missing of one of the electron Fermi pockets near the *B* point in the magnetotransport measurements might due to that its carrier momentum relaxation time is not sufficiently long. The two fundamental frequencies correspond to the external cross-sectional areas of the FS $F_\alpha$ and $F_\beta$ as $A$ = 0.95 and 1.67 nm$^{-2}$, respectively, which are calculated by using the Onsager relation $F = (\hbar/2\pi e)A$. The effective cyclotron mass $m^*$ at $E_F$ can be estimated from the L-K fitting to the $R_T$, as is shown in Fig. 2(e), giving $m^*$ = 0.339 $m_e$ and 0.412 $m_e$ for $F_\alpha$ and $F_\beta$, respectively, where $m_e$ denotes the free electron mass. The Fermi wave vectors for $F_\alpha$ and $F_\beta$ are estimated to be 0.55 and 0.73 nm$^{-1}$ respectively from $k_F = \sqrt{2eF/\hbar}$ and the corresponding Fermi velocities $v_F$ = 1.88×10$^5$ and 2.05×10$^5$ m s$^{-1}$ are calculated from $v_F = \hbar k_F/m^*$. Fitting to the field dependent amplitudes of the quantum oscillations at 1.8 K by using the L-K formula, shown in Fig. 2(f), gives the Dingle temperature $T_D$ = 5.33 and 11.38 K and the corresponding quantum scattering lifetime $\tau_Q$ = 2.28 × 10$^{-13}$ and 1.07× 10$^{-13}$ s, respectively. Furthermore, the quantum mobility $\mu_Q$ ~ 1.2× 10$^3$ and 456.6 cm$^2$ VS$^{-1}$ could also be obtained from $\mu_Q = e\tau_Q/m^*$. The results are summarized in Table I to guide a comprehensive understanding about these parameters.

To achieve in-depth insights into the SdH oscillations, the Landau level (LL) index fan diagram is constructed, aiming to examine the Berry phase of TaSe$_3$ accumulated along the cyclotron orbit. This is due to that the nontrivial Berry phase is generally considered to be a key evidence for Dirac fermions, which are caused by the pseudo-spin rotation under a magnetic field [49, 50]. In the L-K equation, $\gamma$ (= 1/2 − $\phi_B$/2$\pi$) is the Onsager phase factor and $\delta$ represents the FS dimension-dependent correction to the phase shift, which is 0 for 2D system and ± 1/8 for 3D system in presence of nontrivial Berry phase [49,51]. The LL index phase diagram is shown in Fig. 2(d), in which the valley positions of $\Delta\rho_{xx}$ against 1/$B$ were assigned to be integer indices and the peak positions of $\Delta\rho_{xx}$ were assigned to be half-integer indices. It is apparent that all the points nicely fall on a straight line, thus allowing a linear fitting that gives the intercepts of 0.3614 and 0.599 corresponding to $F_\alpha$ and $F_\beta$, respectively, verifying that both Fermi pockets have nonzero Berry phase. The analysis could be further supported by the following theoretical calculations which also suggest topologically nontrivial bands, guaranteeing the reliability of our analysis.

As we noted above, during the longitudinal resistivity $\rho_{xx}$ measurements, $B$ was initially perpendicular to the [1 0 -1] direction, i.e. parallel to the [-1 0 -1] direction. To map out the FS, we rotated the sample by gradually deviating the [-1 0 -1] direction from that of $B$. The rotation geometry is depicted in the inset of Fig. 3(a), in which $\theta$ is the angle between $B$ and the [-1 0 -1] direction. The angle dependent MR is presented in Fig. 3(a), which exhibits clear magnitude change, exposing the anisotropic nature of the band structure. Fig. 3(b) present the angle dependent SdH oscillations with a constant offset in the vertical coordinate, which display clear shift of the peaks with the increases of $\theta$. The $\theta$ dependence of fundamental frequencies derived from the SdH oscillations are shown in Fig. 3(c), changing from 100 T and 175 T at $\theta = 90°$ (out of plane) to 144 T and 216 T at $\theta = 0°$ (in plane), thus unambiguously unveiling the anisotropy of the Fermi pockets associated with the SdH oscillations. Due to the qusi-1D character of the crystal, it is not convenient to measure the in-plane anisotropy by magnetotransport, which therefore leaves a space for further optimization of the crystal. In the *ab initio* calculations, the FSs consist of a large hole pocket enclosing the Γ point and two electron pockets near the $B$ point, as shown in Fig. 3(d). By a careful comparison between theory and experiment, we could reach a conclusion that the $F_\alpha$ comes from one of the electron pockets near the $B$ point, and $F_\beta$ comes from the hole pocket enclosing the Γ point.

More details about the electronic band structure of TaSe$_3$ would be helpful for understanding the magnetotransport properties. The mBJ band structure without spin-orbit coupling (SOC), presented in Fig. 4(a), unveils a semimetal ground state with both electron and hole FSs in TaSe$_3$. The fat-band structure shows that the valence band is dominated by Ta-$d_{z^2}$ orbital while the conduction band is derived from Se-$p_z$ orbital. The valence and conduction bands cross each other along the A-B and D-E lines near the $E_F$, resulting in a band inversion which gives rise to nontrivial band topology of TaSe$_3$. Fig. 4(b) plots the energy bands with SOC, which shows that gaps will be opened at the crossing points, creating a bandgap between valence and conduction bands at each $k$ point. Thus TaSe$_3$ can be considered as an insulator defined on a curved Fermi level between valence and conduction bands. Using the Wilson loop method [52], we calculate the topological invariant $Z_2$. The Wannier charge center (WCC) evolution on the time-reversal invariant planes of TaSe$_3$ is shown in Fig. 4(c), from which we find $Z_2$ = (1;100), indicating that TaSe$_3$ hosts a strong topological insulator state. Due to the nontrivial bulk band topology, protected Dirac cone surface states with spin-momentum locking are expected to appear on the boundary of TaSe$_3$. As shown in Fig. 4(d), the surface states on the (010) surface clearly show a Dirac topological surface states at the $\bar{B}$ point in the bulk band gap. The surface Dirac cone exhibits a helical spin texture, as shown in Fig. 4(e). Our results are nicely in agreement with that in Ref. 34.

Despite of the rather low $T_c$, TaSe$_3$ has many attractive virtues for the study on interplay between nontrivial topology and superconductivity. Unlike FeSe$_{0.45}$Te$_{0.55}$ and (Li$_{0.84}$Fe$_{0.16}$)OHFeSe which need doping to be superconducting, TaSe$_3$ is an intrinsic superconductor hosting topologically nontrivial bands and spin-momentum locking states on the surface. Moreover, high-quality TaSe$_3$ crystal is easy to be grown and optimized, unlike 2M-WS$_2$, another intrinsic topological superconductor hosting Majorana zero mode, which is rather difficult in growth of high-quality crystals [52]. However, it should be very cautious to examine the possible CDW order in TaSe$_3$. It is known that CDW order appears in its analogues such as TaS$_3$ [36] and NbSe$_3$ [54], which whereas had not been detected in TaSe$_3$. The calculations on phonon dispersion of TaSe$_3$ show conflictive results on the imaginary frequency that characterizes the CDW. One claimed that the phonon dispersion shows imaginary frequencies [55] whereas the other one argued that it can be safely excluded [34]. A very recent study on dimension controllable TaSe$_3$ mesowires suggested the observation of CDW by detecting an anomaly in the resistivity at 65 K [56]. The emergency of CDW order is generally related to the FS folding in the electronic structure, which may influence the nontrivial band structure. These conflicts definitely appeal further investigations to achieve a clarification.

## IV. CONCLUSION

In summary, by performing high magnetic field magnetotransport measurements and *ab initio* calculations on the layered superconductor $TaSe_3$, we have observed topologically nontrivial electronic bands stemming from the three-dimensional strong topological insulator state. We also have unveiled spin-momentum locking surface states. The results demonstrate the intrinsic superconductor $TaSe_3$ as an ideal platform for the study of the interplay between topological phase and superconductivity. However, to realize Majorana Fermion, the surface states, which strongly depends on the position of chemical potential, should be well separated from the bulk states. If the chemical potential lies ~ 40 meV below the $E_F$, as is shown in Fig. 4(d), the surface states will merge into the bulk states. In such case, fine controlling the position of chemical potential through quantum manipulation, such as chemical doping, doping carriers from gating in a Field effect transistor configuration, applying external pressure or strain, is definitely requisite.

**ACKNOWLEDEMENTS**

The authors acknowledge the support by the Natural Science Foundation of Shanghai (Grant No. 17ZR1443300), the National Natural Science Foundation of China (Grant No. 11874264, 51772145), the strategic Priority Research Program of Chinese Academy of Sciences (Grant No. XDA18000000), and the Key Scientific Research Projects of Higher Institutions in Henan Province (19A140018). Y.F.G. acknowledges the starting grant of ShanghaiTech University and the Program for Professor of Special Appointment (Shanghai Eastern Scholar). W. Z. is supported by the Shenzhen Peacock Team Plan (Grant No. KQTD20170809110344233), and Bureau of Industry and Information Technology of Shenzhen through the Graphene Manufacturing

Innovation Center (Grant No. 201901161514). Y.F.H acknowledges Distinguished Young Scholars Fund of Jiangsu Province (BK20180003) and JiangHai Talent program of Nantong.

TABLE I. Parameters derived from SdH oscillation for TaSe$_3$.

| $F$ (T) | $A$ (nm$^{-2}$) | $K_F$ (nm$^{-1}$) | $v_F$ (m/s) | $E_F$ (meV) | $m^*/m_e$ | $T_D$ (K) | $\tau_Q$ (s) | $\mu_Q$ (cm$^2$/Vs) | Berry phase |
|---|---|---|---|---|---|---|---|---|---|
| 100 | 0.95 | 0.55 | $1.88 \times 10^5$ | 68.16 | 0.339 | 5.33 | $2.28 \times 10^{-13}$ | 1182.5 | $0.72\pi$ |
| 175 | 1.67 | 0.73 | $2.05 \times 10^5$ | 98.49 | 0.412 | 11.38 | $1.07 \times 10^{-13}$ | 456.6 | $1.20\pi$ |

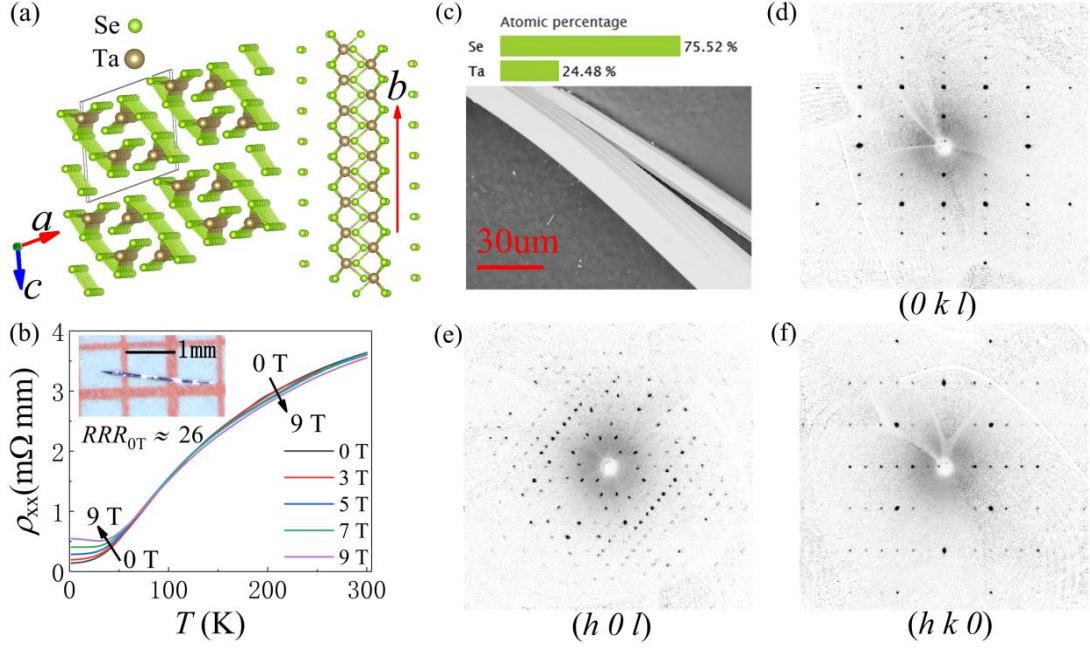

**Fig. 1.** (Color Online) (a) Schematic crystal structure of TaSe$_3$. (b) Temperature dependence of the longitudinal resistivity $\rho_{xx}$ at different magnetic fields. Inset shows a picture of a typical crystal. (c) The stoichiometry of TaSe$_3$ crystal measured by the EDS spectrum. (d) - (f) Diffraction patterns in the reciprocal space along (*0 k l*), (*h 0 l*), and (*h k 0*) directions.

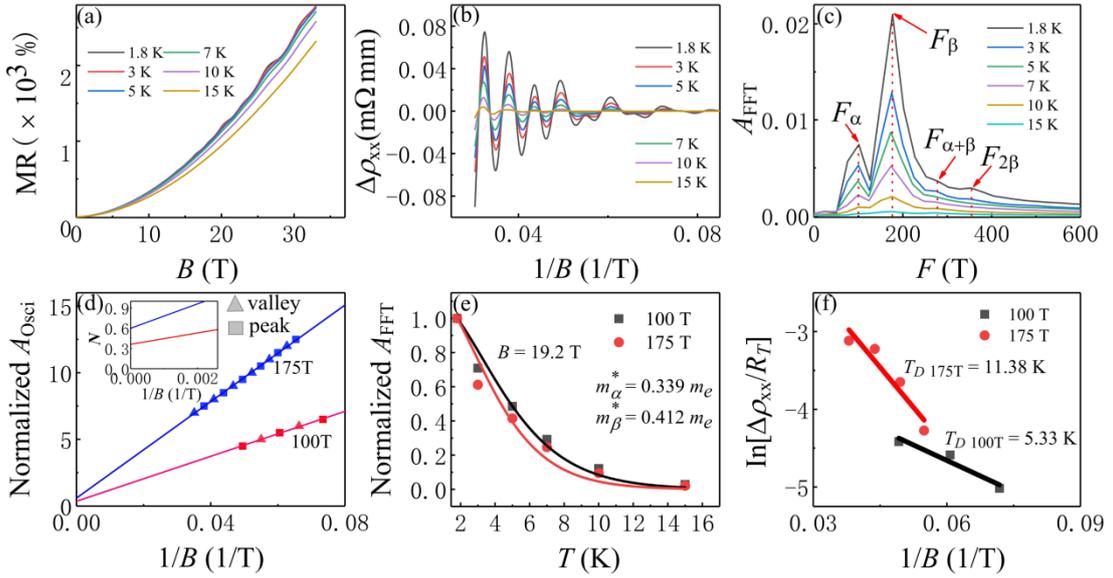

**Fig. 2.** (Color Online) (a) Magnetic field dependent MR at different temperatures below 15 K, exhibiting quantum oscillations when $B > 12$ T. (b) SdH oscillatory component as a function of $1/B$ after subtracting the smooth background of MR. (c) FFT spectra of $\Delta\rho_{xx}$. (d) Landau index $N$ plotted against $1/B$ at various $\theta$ at 1.5 K. Inset enlarges the intercepts of the fitting. Triangles

represent the valley positions and squares represent the peak positions of the SdH oscillations. (e) Temperature dependence of relative FFT amplitudes of the SdH oscillations. The solid lines denote the fitting by using the L-K formula. (f) Dingle plot of the SdH oscillation at $T$ = 1.8 K.

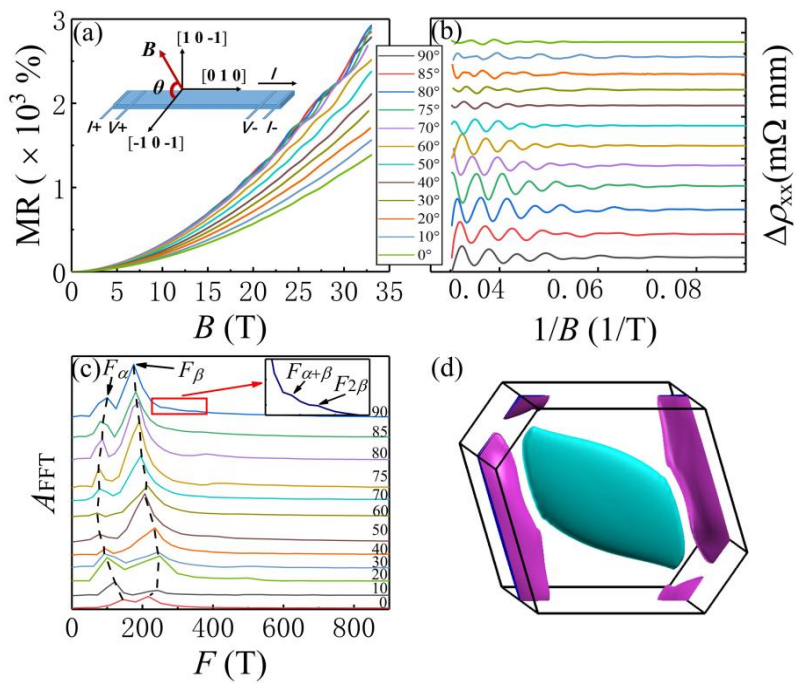

**Fig. 3.** (Color Online) (a) Longitudinal MR versus magnetic field $B$ at various $\theta$ when $T$ = 1.8 K. Inset shows the schematic measurement configuration. (b) SdH oscillatory component as a function of $1/B$ at different angles. (c) FFT spectra of $\Delta\rho_{xx}$ at various $\theta$. (d) The Fermi surface of TaSe$_3$.

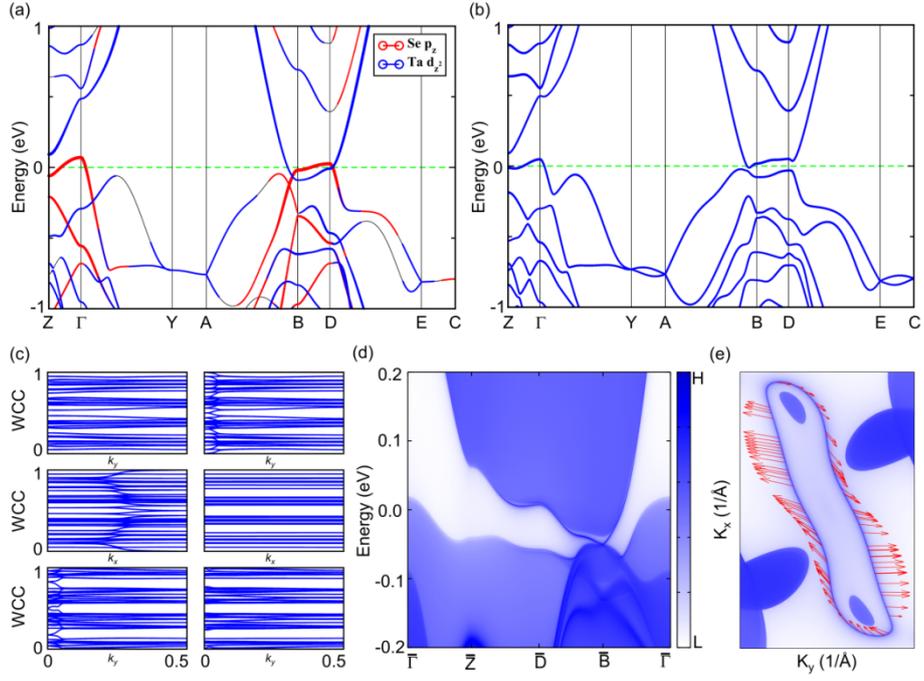

**Fig. 4.** (Color Online) Electronic band structure of TaSe$_3$ (a) without and (b) with SOC considered at mBJ level. The symbol size in (a) corresponds to the projected weight of Bloch states onto the Ta-$d_{z^2}$ (blue) and Se-$p_z$ (red) orbits. (c) Wannier charge center evolution on the time-reversal invariant planes of TaSe$_3$. (d) Calculated surface states of TaSe$_3$ on the (010) surface. (e) Fermi surface and corresponding spin texture at a fixed energy $E - E_F$ = -40 meV of the topological surface states in (d).